\documentclass[prl,superscriptaddress,amsmath,amssymb,aps,twocolumn,showpacs,floatfix,longbibliography]{revtex4-2}

\usepackage{amsmath}
\usepackage{inputenc}
\usepackage{graphicx}
\usepackage{dcolumn}
\usepackage{bm}
\usepackage{xcolor}
\usepackage{hyperref}

\begin{document}

\title{Bragg condition for scattering into a guided optical mode}

\author{B. Olmos}
\affiliation{School of Physics and Astronomy and Centre for the Mathematics and Theoretical Physics of Quantum Non-Equilibrium Systems, The University of Nottingham, Nottingham, NG7 2RD, United Kingdom}
\affiliation{Institut f\"ur Theoretische Physik, Universit\"at T\"ubingen, Auf der Morgenstelle 14, 72076 T\"ubingen, Germany}
\author{C. Liedl}
\affiliation{Department of Physics, Humboldt-Universit\"at zu Berlin, 10099 Berlin, Germany}
\author{I. Lesanovsky}
\affiliation{School of Physics and Astronomy and Centre for the Mathematics and Theoretical Physics of Quantum Non-Equilibrium Systems, The University of Nottingham, Nottingham, NG7 2RD, United Kingdom}
\affiliation{Institut f\"ur Theoretische Physik, Universit\"at T\"ubingen, Auf der Morgenstelle 14, 72076 T\"ubingen, Germany}
\author{P. Schneeweiss}
\affiliation{Department of Physics, Humboldt-Universit\"at zu Berlin, 10099 Berlin, Germany}

\date{\today}

\begin{abstract}
We theoretically investigate light scattering from an array of atoms into the guided modes of a waveguide. We show that the scattering of a plane wave laser field into the waveguide modes is dramatically enhanced for angles that deviate from the geometric Bragg angle. We derive a modified Bragg condition, and show that it arises from the dispersive interactions between the guided light and the atoms. Moreover, we identify various parameter regimes in which the scattering rate features a qualitatively different dependence on the atom number, such as linear, quadratic, oscillatory or constant behavior. We show that our findings are robust against voids in the atomic array, facilitating their experimental observation and potential applications. Our work sheds new light on collective light scattering and the interplay between geometry and interaction effects, with implications reaching beyond the optical domain.
\end{abstract}

\maketitle

\textit{Introduction.} Bragg diffraction was originally discovered when investigating crystalline solids using X-rays. However, Bragg scattering is based on the constructive interference of partial waves that originate from periodically arranged scatterers and is, thus, a very general phenomenon that plays a central role in many branches of physics, most notably in optics~\cite{AshcroftBook}. One well-known and technologically relevant application of Bragg scattering are dielectric mirrors, which enable the reflection of light without almost any losses. More recently, Bragg scattering phenomena that occur when laser-cooled atoms are used as scatterers for light have been the matter of numerous theoretical and experimental studies~\cite{Stenger_1999,Zoubi_2010,Weitenberg_2011,Chang_2012,Kornovan_2016,Corzo_2016,Soerensen_2016,Olmos_2020,Meng_2020}.

While the resonances of the materials the dielectric mirror is made of are far-detuned with respect to the wavelength of the reflected light, this can be distinctly different in the case of atomic scatterers. When the light is resonant or near-resonant with an atomic transition, the light can be absorbed by the atom, with the scattered light acquiring a phase shift relative to the incident light. Close to resonance, the scattering cross section is significantly enhanced, such that multiple scattering between different atoms becomes relevant~\cite{Kaiser_2009,Sokolov_2019}. Moreover, single atoms can scatter only one photon at a time, giving rise to non-linear optical effects~\cite{Kolchin_2011,Chang_2014,Firstenberg_2016}. The interplay between Bragg scattering and cooperative effects stemming from coherent scattering of light between emitters gives rise to surprising phenomena, such as photonic band gaps~\cite{LeKien_2014}, sub-radiant atomic mirrors~\cite{Bettles_2016,Rui_2020}, improved optical quantum memories~\cite{Facchinetti_2016,Asenjo_2017}, guided light in atomic chains \cite{Asenjo_2019,Masson_2020}, collective enhancement of chiral photon emission into a waveguide~\cite{Jones_2020}, or the modification of Bragg scattering from atoms in an optical lattice \cite{Birkl_1995}.  

\begin{figure}
\begin{center}
    \includegraphics[width=\columnwidth]{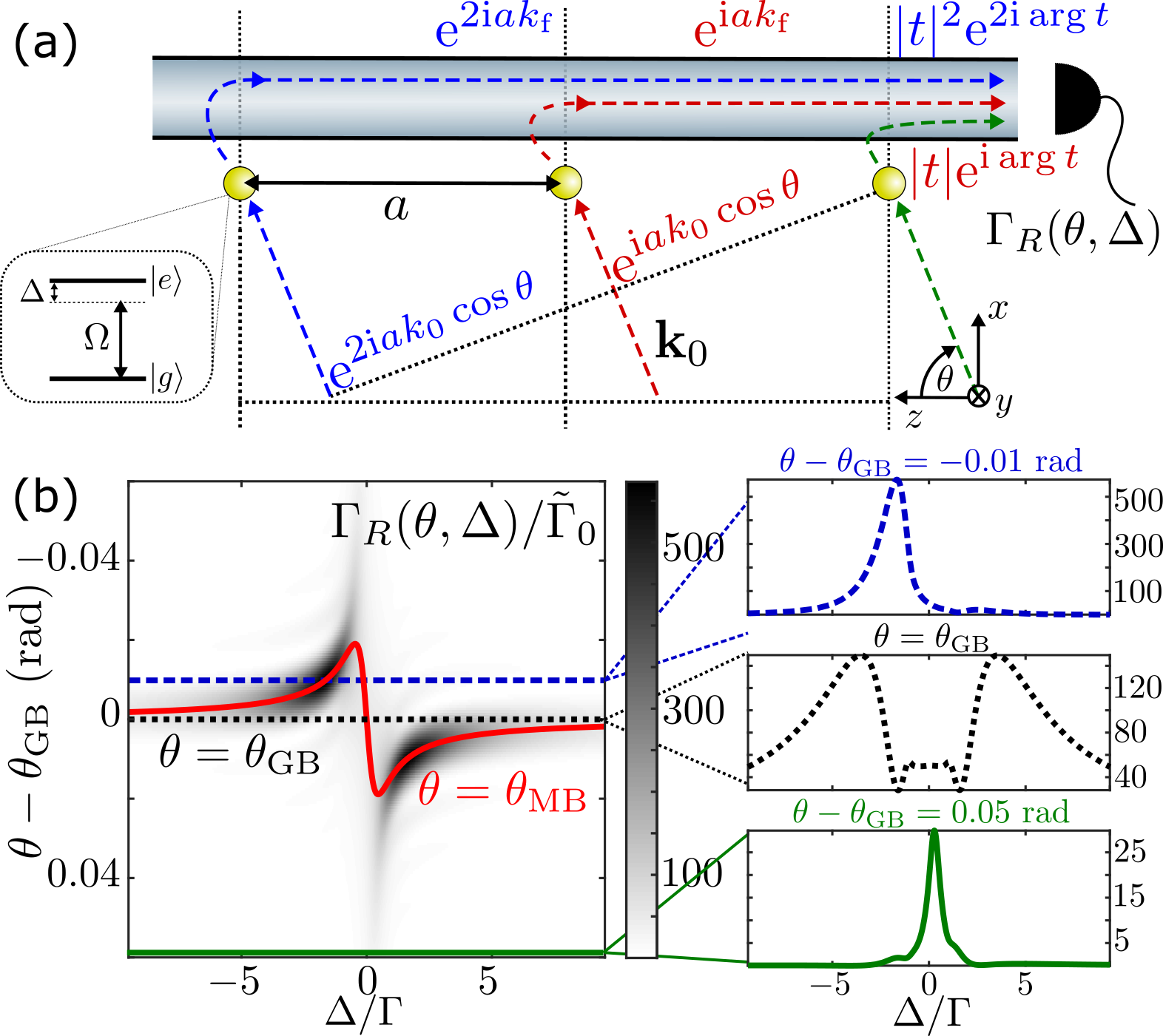}
    \caption{\textbf{Scattering into waveguide mode.} (a): An array of emitters coupled to a waveguide is driven by a laser with wave vector $\mathbf{k}_0$ (forming an angle $\theta$ with the array), Rabi frequency $\Omega$, and detuning $\Delta$. 
    The rate of photons emitted into the guided mode propagating to the right, $\Gamma_R(\theta,\Delta)$, can be well approximated by considering the interference of the scattering processes indicated with the dashed, colored arrows. (b): $\Gamma_R(\theta,\Delta)/\tilde{\Gamma}_0$, where $\tilde{\Gamma}_0$ is the single-atom scattering rate into the waveguide on resonance. The maxima of the scattering rate occur at an angle $\theta_\mathrm{MB}$ (red line) that deviates from the geometric Bragg angle $\theta_\mathrm{GB}$, and $\Delta\neq 0$. The cuts on the right show qualitatively different spectra depending on the choice of $\theta$. Here, $N=144$ and $D=1$.}
    \label{fig:Schematic}
\end{center}
\end{figure}

In this work we theoretically investigate the scattering of light from an atomic emitter array into the guided optical modes of a waveguide. The emitters are coherently driven by an external plane wave light field such that the scattered light from the different emitters can interfere constructively. We demonstrate that the dispersive waveguide-mediated atom-atom interactions lead to a modified Bragg condition, i.e., the maximum scattering rate into the guided mode is reached at laser incidence angles different from the one determined by the geometric Bragg relation. Here the maximum scattering rate is shown to be dramatically enhanced and to grow linearly with the number of emitters. This is in stark contrast to other incidence angles for which a saturation is observed. We also identify situations in which the scattering rate scales quadratically and even oscillates as a function of the atom number. Strikingly, all these qualitatively different scalings are shown to be largely independent of the asymmetry (or ``chirality'') of the emitter-waveguide coupling \cite{Lodahl_2017} and also robust against voids in the atomic array. 

\textit{System.} We consider a one-dimensional array of $N$ atomic emitters with nearest neighbor distance $a$ situated parallel to an optical waveguide (here a silica nanofiber), as sketched in Figure \ref{fig:Schematic}(a). Each emitter is modelled as a two-level system with ground and excited states denoted by $\left|g\right>$ and $\left|e\right>$, respectively. 
The atoms are externally driven by a plane wave monochromatic light field with Rabi frequency $\Omega$, detuning $\Delta$, and wave vector $\mathbf{k}_0$ that encloses an angle $\theta$ with the array. When an atom is excited, it can decay back into its ground state emitting a photon with wavelength $\lambda=2\pi/k_0$ with $k_0=|\mathbf{k}_0|$. Due to the proximity of the nanofiber, the photon can be emitted into one of the two counter-propagating guided modes supported by the nanofiber (at a rate $\gamma_R$ and $\gamma_L$ for the right- and left-propagating mode, respectively). It also can be emitted into the unguided modes (at a rate $\gamma_\mathrm{u}$), whose modification due to the presence of the fiber is taken into account \cite{LeKien_2017,Asenjo_2017b}. The efficiency of the coupling into the guided modes is quantified by the so-called beta factor $\beta=(\gamma_R+\gamma_L)/\Gamma$, where $\Gamma=\gamma_R+\gamma_L+\gamma_\mathrm{u}$ is the total single-atom decay rate. Moreover, depending on the orientation of the dipole moment of the atomic transition, an asymmetry of the emission into the guided modes can be present, such that $\gamma_R\neq\gamma_L$~\cite{Lodahl_2017}. We will quantify this asymmetry via the parameter $D=(\gamma_R-\gamma_L)/(\gamma_R+\gamma_L)$.

Under the Born-Markov approximation, the dynamics and stationary state of the system are determined by the master equation \cite{Asenjo_2017b,LeKien_2017}
\begin{eqnarray}\label{eq:master}
    \dot{\rho}&=&-\frac{\mathrm{i}}{\hbar}\left[H_\mathrm{L},\rho\right]-\mathrm{i}\sum_{j\neq l}\left[V_{jl}\sigma^\dag_j\sigma_l,\rho\right]\\\nonumber
    &&+\sum_{jl}\Gamma_{jl}\left(\sigma_j\rho\sigma_l^\dag-\frac{1}{2}\left\{\sigma_j^\dag\sigma_l,\rho\right\}\right),
\end{eqnarray}
where $\sigma_j=\left|g_j\right>\left<e_j\right|$ for $j=1,\dots N$. Here, the first term describes the action of the laser field:
\begin{equation}
    H_\mathrm{L}=\hbar\sum_{j=1}^N \left[\Omega\left(\mathrm{e}^{\mathrm{i}k_0{aj}\cos{\theta}}\sigma_j^\dag+\mathrm{h.c.}\right)+\Delta \sigma_j^\dag\sigma_j\right].
\end{equation}
Note, that in the following we will assume that the laser driving is weak, such that the saturation parameter is small, i.e. $\Omega\ll\Gamma$. The second term in Eq. (\ref{eq:master}) represents dipole-dipole interactions induced by the exchange of virtual photons between the $j$-th and $l$-th atom at a rate $V_{jl}$. Finally, the last term describes the incoherent emission of photons. The decay rates of the entire system of coupled atoms, $\gamma_c$, are given by the eigenvalues of the dissipation coefficient matrix $\Gamma_{jl}$. For a single atom the decay rate is simply $\Gamma$, i.e., the atom's excited state population decay rate including possible modification due to the proximity of the nanofiber \cite{LeKien_2005,Scheel_2015}. For several atoms, the decay becomes collective, and the corresponding decay rates $\gamma_c$ can be either superradiant ($\gamma_c\gg\Gamma$), or subradiant ($\gamma_c\ll\Gamma$) \cite{Solano_2017b,Jones_2020,Zhang_2020}.

It is convenient to separate the contribution of the guided 
and unguided modes 
in both the coherent and incoherent interaction matrix coefficients as $V_{jl}= V_{jl}^{R}+V_{jl}^{L}+V_{jl}^\mathrm{u}$ and $\Gamma_{jl}=\Gamma_{jl}^{R}+\Gamma_{jl}^{L}+\Gamma_{jl}^\mathrm{u}$, respectively. The character of the interactions mediated by the unguided modes is fundamentally different from that of the guided ones: while the unguided modes give rise to interactions that decay with the distance between the atoms, the interactions mediated by guided modes are infinite-ranged \cite{Ramos_2014,Pichler_2015,LeKien_2017,Asenjo_2017b}. 

We are here interested in the photon emission rate into the guided modes. In particular, we will analyze the scattering rate into the right-propagating guided mode in the stationary state, defined as
\begin{equation}\label{eq:GammaR}
\Gamma_R (\theta, \Delta)=\sum_{jl}\Gamma^R_{jl}\left<\sigma_j^\dagger\sigma_l\right>_\mathrm{ss}~.
\end{equation}
As one can observe in Fig.~\ref{fig:Schematic}(b), the resonances of the fluorescence excitation spectrum $\Gamma_R(\theta,\Delta)$ are qualitatively modified when the angle $\theta$ is close to the one given by the conventional, geometric Bragg condition. The latter is given by $\cos{\theta_\mathrm{GB}}=2\pi m/(ak_0)-k_\mathrm{f}/k_0$, with $m \in \mathbf{Z}$ and $k_\mathrm{f}$ being the propagation constant inside the nanofiber. For most angles the spectrum is well approximated by a Lorentzian centered at $\Delta=0$. As $\theta_\mathrm{GB}$ is approached, the scattering rate increases rapidly. The spectrum then starts to display a maximum that is off resonance, in particular for an excitation under the modified Bragg angle $\theta_\mathrm{MB}$. Exactly at $\theta=\theta_\mathrm{GB}$, the spectrum splits symmetrically around $\Delta=0$ into two peaks \cite{Olmos_2020}.

\textit{Scattering into the waveguide for unidirectional coupling.} In order to understand the origin of the intricacies of the spectrum and to investigate the scaling with the system parameters, we make use of a simplified model which reproduces the main features found with the full one described by Eq. (\ref{eq:master}). In this model, we only account for the waveguide-mediated interactions. The decay into the unguided modes is considered to be diagonal: each atom decays with rate $\gamma_\mathrm{u}$ into the unguided modes and no interactions are induced between the atoms via this dissipative channel. Moreover, for simplicity we consider that the coupling into the waveguide is fully directional, i.e., $\gamma_L=0$ and $D=1$. Here, the scattering rate (\ref{eq:GammaR}) is found to be
\begin{equation}\label{eq:gamma_chiral}
    \Gamma_R(\theta,\Delta)=\tilde{\Gamma}_\Delta\left|\sum_{m=0}^{N-1}t^m e^{-\mathrm{i}m k_\mathrm{eff}a}\right|^2,
\end{equation}
where we have defined the effective wave number $k_\mathrm{eff}=k_0\cos{\theta}+k_\mathrm{f}$, the single-atom scattering rate into the guided modes $\tilde{\Gamma}_\Delta=4\Omega^2\beta\Gamma/(4\Delta^2+\Gamma^2)$, and the complex-valued amplitude transmission coefficient $t=1-2\mathrm{i}\beta\Gamma/(2\Delta+\mathrm{i}\Gamma)$.

Expression (\ref{eq:gamma_chiral}) can be alternatively obtained by coherently summing the contributions of the light scattered into the waveguide by all atoms, as illustrated in Fig. \ref{fig:Schematic}(a). The light that each atom scatters into the waveguide exhibits a phase difference with respect to the one emitted by an atom that is one lattice constant to its right: The contribution related to the plane wave excitation is given by $k_0 a \cos{\theta}$ and the one due to propagation in the fiber is $k_\mathrm{f}a$. Moreover, $t$ describes the transmission of the light when it passes an atom in the chain, yielding a phase shift given by $\arg{t}$ and an amplitude reduction $|t|$. The total scattering rate $\Gamma_R$ is then obtained as the absolute value squared of the sum of all amplitudes, multiplied by $\tilde{\Gamma}_\Delta$.

\begin{figure}[t!]
    \includegraphics[width=\columnwidth]{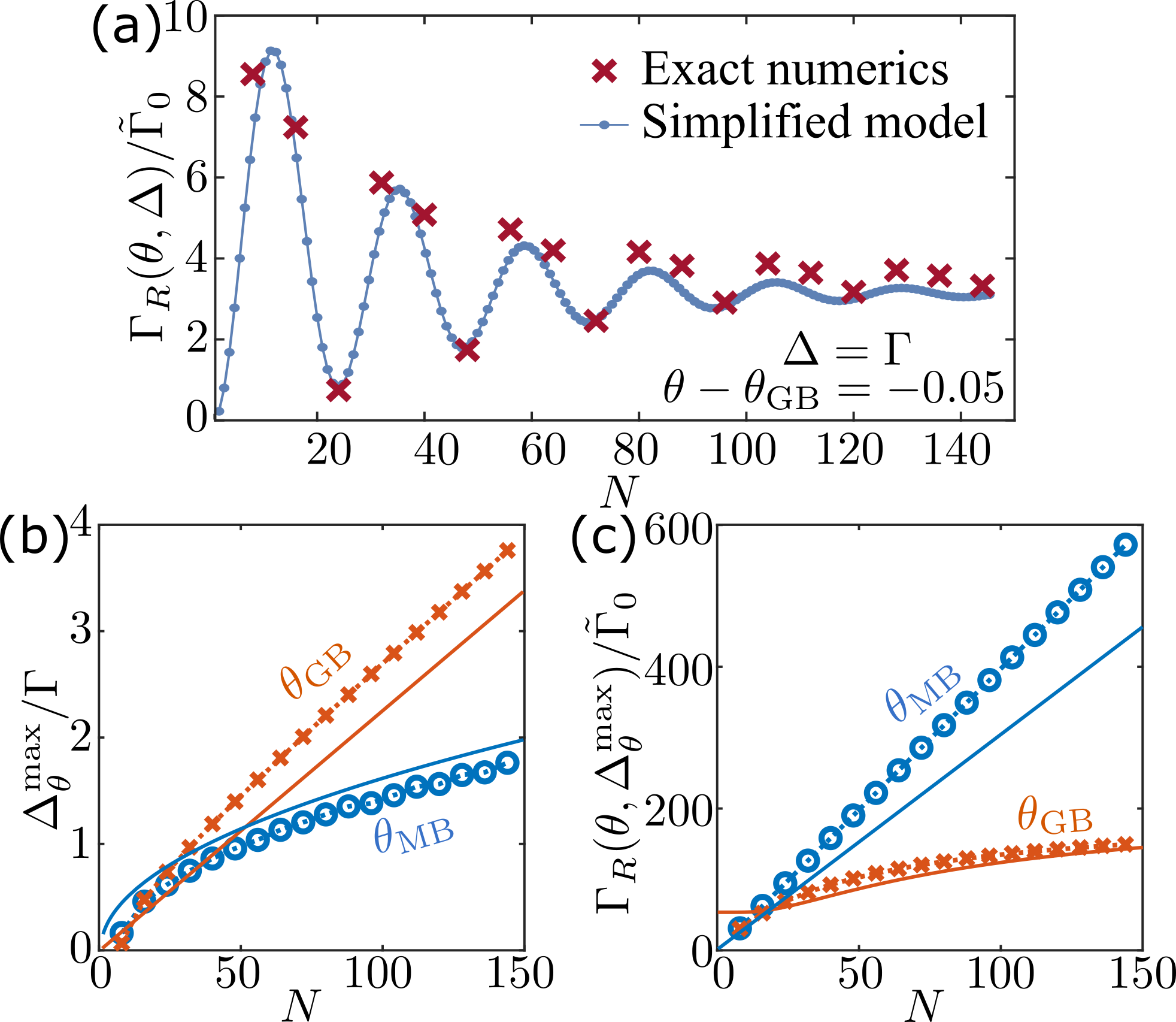}
    \caption{\textbf{Scattering rate for unidirectional coupling.} (a): Normalized scattering rate into the right-propagating  mode for  $\theta-\theta_\mathrm{GB}=-0.05$ rad and $\Delta=\Gamma$ as a function of $N$. (b): Detuning and (c): maximum scattering rate as a function of $N$ as the laser drives the chain at $\theta_\mathrm{GB}$ (red) and $\theta_\mathrm{MB}$ (blue). In all cases, (a)-(c), there is a good agreement between the predictions from the simplified model (lines) and the full master equation (markers).}
    \label{fig:At_mod_Bragg}
\end{figure}

In order to analyze the dependence of the scattering rate with the number of atoms $N$, we perform the sum in (\ref{eq:gamma_chiral}) formally such that
\begin{equation}\label{eq:gamma_general}
    \Gamma_R(\theta,\Delta)=\tilde{\Gamma}_\Delta\,\frac{1 + |t|^{2N}- 2 |t|^N \cos(b N)}{1 + |t|^{2}- 2 |t| \cos(b)},
\end{equation}
where $b=\arg{t}-k_\mathrm{eff}a$. The numerator contains a term due to which the scattering rate oscillates as a function of $N$, see Fig.~\ref{fig:At_mod_Bragg}(a) for an example. Since $N$ can only take integer values, the oscillations are sampled with a frequency $f_s=1$, and one observes oscillations at an angular aliasing frequency $b_{\text{alias}}=\text{min}||b|-2\pi kf_s|$, for $k\in\mathbb{N}_0$. 
From expression (\ref{eq:gamma_general}), one can also see that the oscillations are damped via the term $|t|^N$ describing the field decay, such that for large enough $N$ a saturation value is reached. Conversely, for small values of $N$, the scattering rate grows proportionally to $N^2$ for all $|N\ln{|t|}|\ll1$ and provided that $N$ is smaller than the period of the oscillations, $N<N_p=2\pi/b_\mathrm{alias}$. 

We now discuss the case when the atoms are driven under the geometric Bragg condition~\cite{Jones_2020}, i.e. $\theta=\theta_\mathrm{GB}$. Here, $b=\arg{t}-2m\pi$, and, for large enough atom numbers $N$, the spectrum splits into two peaks [see Fig.~\ref{fig:Schematic}(b)]. 
In the limit $\Delta\gg\Gamma$, and for large $N$, we find that the detunings at which the two maxima occur are approximately given by
\begin{equation}\label{eq:Deltamax}
    \Delta_{\theta_\mathrm{GB}}^\mathrm{max}\approx \pm\frac{\Gamma\beta N}{\pi},
\end{equation}
with corresponding maximum scattering rate
\begin{equation}\label{eq:Gammamax}
    \Gamma_R(\theta_\mathrm{GB},\Delta^\mathrm{max}_{\theta_\mathrm{GB}})\approx\frac{\Omega^2}{\beta\Gamma}\left(1 + \mathrm{e}^{-\frac{\pi^2(1-\beta)}{2\beta N}}\right)^2.
\end{equation}
Hence, the maximum scattering rate approaches a saturation value $4\Omega^2/(\beta \Gamma)$ when $N\to\infty$. Note that, rather counterintuitively, this saturation value is larger the weaker the coupling $\beta$.

Eq. (\ref{eq:gamma_general}) allows to infer a \emph{modified} Bragg condition
\begin{equation}\label{eq:mod}
    \cos{\theta_\mathrm{MB}}=\cos{\theta_\mathrm{GB}}+\frac{\arg{t}}{k_0 a}.
\end{equation}
Here $b$ is a multiple of $2\pi$, the scattering rate reaches a maximum and it does not oscillate with $N$. In agreement with the numerical results shown in Fig.~\ref{fig:Schematic}(b), the maximum guided scattering rate is therefore not assumed when the emitter array is driven at the geometric Bragg angle, but rather slightly away from it. As $\arg{t}$ depends on the detuning, so does $\theta_\mathrm{MB}$, as depicted by the red solid line in Fig.~\ref{fig:Schematic}(b). 
As can also be seen, the maxima of the spectrum $\Gamma_R(\theta_\mathrm{MB},\Delta)$ are shifted away from resonance. However, comparing to the geometric Bragg condition case, we find different scalings of the optimal detunings and the maximum scattering rate with the number of atoms $N$, given approximately by
\begin{equation}\label{eq:Delta_max_mod}
    \Delta^\mathrm{max}_{\theta_\mathrm{MB}}\propto\pm\sqrt{N (1-\beta) \beta}\Gamma,
\end{equation}
and
\begin{equation}\label{eq:Gamma_max_mod}
    \Gamma_{R,\mathrm{MB}}^\mathrm{max}\propto\frac{\Omega^2 N}{(1-\beta)\Gamma},
\end{equation}
respectively. Notably, now the maximum scattering rate does not saturate for large values of $N$, but rather grows linearly with $N$, eventually diverging as $N\to\infty$. As a consequence, while there is a collective enhancement of the total scattering for excitation under $\theta_\mathrm{GB}$, the scattering rates are dramatically further enhanced at $\theta_\mathrm{MB}$. For example, in the case shown in Fig. \ref{fig:At_mod_Bragg}(c), 150 atoms can scatter as much as $\sim$600 independent atoms into the waveguide. Finally, note that all the discussed scalings are confirmed by the numerical simulation of the full master equation (\ref{eq:master}), cf.~Fig.~\ref{fig:At_mod_Bragg}.

\begin{figure}[t!]
    \includegraphics[width=\columnwidth]{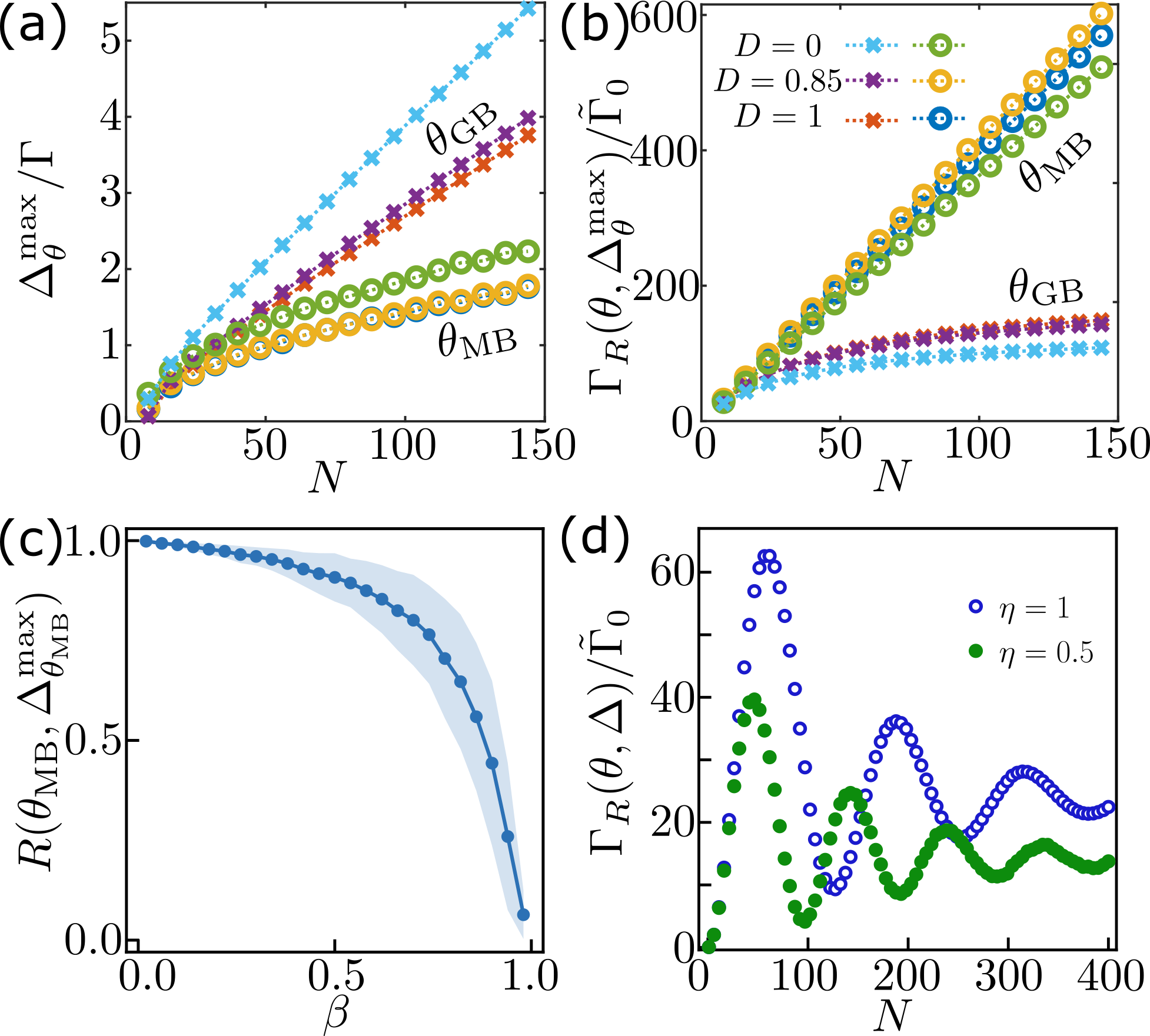}
    \caption{\textbf{Asymmetry and voids.} (a): Detuning $\Delta^\mathrm{max}_\theta$ and (b): maximum scattering rate, as a function of $N$ as the laser drives the chain at the geometric Bragg angle (GB) and the modified (MB) one, by numerical evaluation of the full master equation \eqref{eq:master}. Data is shown for $D=0,0.85$ and $1$, and $\gamma_R/\Gamma=0.0707$. (c): 
    Robustness $R(\theta_{\mathrm{MB}},\Delta^{\mathrm{max}}_{\theta_\mathrm{MB}}) =\Gamma_R^{\mathrm{voids}}(\theta_{\mathrm{MB}},\Delta^{\mathrm{max}}_{\theta_\mathrm{MB}})/\Gamma_R(\theta_{\mathrm{MB}},\Delta^{\mathrm{max}}_{\theta_\mathrm{MB}})$ against random voids with a filling factor of $\eta=0.5$. We average over $1000$ randomly chosen void configurations and display the standard deviation of the scattering rate as the shaded area. 
    (d): Oscillations of the scattering rate with $N$ for $\theta=\theta_\mathrm{GB}+0.004$ rad and $\Delta=-2\Gamma$. When random voids (green filled circles, averaged over $1000$ realizations) are introduced, the contrast decreases compared to the perfectly filled chain (blue empty circles).}
    \label{fig:robust}
\end{figure}

\textit{Asymmetric and symmetric coupling.} Up to now, we have assumed the special situation where the emission into the guided modes is completely unidirectional, i.e. $D=1$. While this allowed us to obtain analytic results, this is usually not the situation found in realistic experimental settings, where $|D|<1$ or even $D=0$, i.e.~there is symmetric emitter-waveguide coupling. 
We have investigated this situation numerically and found that the scaling with $N$ 
is independent of the value of $D$. This is exemplified in Fig.~\ref{fig:robust}(a) and (b), where we compare the scaling of $\Delta^\mathrm{max}_{\theta}$ and $\Gamma_R(\theta,\Delta^\mathrm{max}_{\theta})$ obtained for $\theta=\theta_\mathrm{MB}$ and $\theta_\mathrm{GB}$ for different values of $D$.

\textit{Robustness against voids.} In experiments, laser-cooled atoms can be trapped next to an optical waveguide in a periodic array of trapping sites \cite{Vetsch_2010}. Here, while the residual motion of the atoms is small enough to observe Bragg scattering phenomena \cite{Corzo_2016,Soerensen_2016}, it is challenging to obtain atomic arrays where indeed every trapping site is occupied. We investigate the robustness of our findings against voids in the atomic array using our simplified model. For this purpose, we simulate an array of $N_\mathrm{sites}=N/\eta$ sites, with $\eta$ being the filling factor.

In Fig.~\ref{fig:robust}(c), we show the average scattering rate over $1000$ randomly chosen configurations for $N=50$ atoms distributed over $N_\mathrm{sites}=100$ sites ($\eta=0.5$, as e.g.~in Ref.~\cite{Vetsch_2012}). For this filling factor, the average spacing between two atoms is $2a$. Hence, here we use the modified Bragg angle \eqref{eq:mod} corresponding to a lattice constant of $2a$ in order to maximize the scattering rate, evaluated at $\Delta=\Delta^\mathrm{max}_{\theta_\mathrm{MB}}$. We compare this scattering rate with the maximum value that is obtained for a completely filled array with $50$ atoms and nearest neighbor distance $a$ as a function of the atom--waveguide coupling. For small $\beta$, the scattering rate for arrays with $\eta=0.5$ is almost as high as for perfect filling, confirming the robustness against voids. However, as $\beta\rightarrow 1$ the robustness decreases which can be understood as follows: At $\theta_{\text{MB}}$, each void leads to a phase difference of $-k_{\text{eff}}a=2m\pi-\arg{t}$ compared to the perfect chain. For small $\beta$, $\arg{t}$ is also small, such that the phase shift due to a void is close to a multiple of $2\pi$ and therefore the scattering properties are not significantly altered. 
For $\beta\rightarrow 1$ however, $\arg{t} \rightarrow \pi$ and thus the voids inhibit the build-up of constructive interference along the chain.

In Fig.~\ref{fig:robust}(d), we study the influence of imperfect filling on the oscillations of the scattering rate with $N$ at an angle $\theta$ slightly away from the Bragg resonance $\theta_\mathrm{GB}$. 
One can see that, despite the imperfect filling, the oscillations are still visible, although they feature a smaller amplitude and a larger oscillation frequency. For an arbitrary filling factor $\eta$, this frequency is simply given by $b_{\text{voids}}=\arg{t}-k_{\text{eff}}\frac{a}{\eta}$. 
These findings indicate that an observation of the effects presented in this work are within reach of current experimental capabilities.

\textit{Conclusion and outlook.} We have studied the collective emission of an array of atoms into a single guided optical mode upon excitation with a plane wave. We show that waveguide-mediated atom-atom interactions lead to a qualitative modification of the Bragg scattering condition. We find simple analytical expressions for the scattering rate into the waveguide and reveal four regimes, each one exhibiting a different scaling with the number of emitters. These findings are shown to be robust against changes in the asymmetry of the coupling and also against voids in the emitter array.

We have first indications that not only the scattering into the guided mode studied here but also the total scattering of a waveguide-coupled array shows collective effects, leading for example to a stronger extinction of the excitation light field compared to a free-space atomic array. Moreover, we noticed that the emission spectrum of the coupled emitters into unguided modes can be perfectly spectrally flat over a large range of detunings, despite the fact that each individual emitter has a Lorenzian line shape \cite{LeKien_2014}. In addition to further investigating these observations, future work will include studying non-linear effects \cite{Kolchin_2011,Masson_2020}, the generalization of these results for other scatterers such as plasmonic nanostructures \cite{Kelf_2005}, and the exploitation of the described effects for quantum information transfer.

\begin{acknowledgments}
We thank A.~Rauschenbeutel and J.~Volz for insightful comments and discussions. Financial support from the European Union's Horizon 2020 research and innovation program under grant agreement No.~800942 (ErBeStA) is gratefully acknowledged. We also acknowledge funding by the Alexander von Humboldt Foundation in the framework of the Alexander von Humboldt Professorship endowed by the Federal Ministry of Education and Research. BO was supported by the Royal Society and EPSRC [Grant No. DH130145]. IL acknowledges support from the “Wissenschaftler-R\"uckkehrprogramm GSO/CZS” of the Carl-Zeiss-Stiftung and the German Scholars Organization e.V.
\end{acknowledgments}


\begin{thebibliography}{36}%
\makeatletter
\providecommand \@ifxundefined [1]{%
 \@ifx{#1\undefined}
}%
\providecommand \@ifnum [1]{%
 \ifnum #1\expandafter \@firstoftwo
 \else \expandafter \@secondoftwo
 \fi
}%
\providecommand \@ifx [1]{%
 \ifx #1\expandafter \@firstoftwo
 \else \expandafter \@secondoftwo
 \fi
}%
\providecommand \natexlab [1]{#1}%
\providecommand \enquote  [1]{``#1''}%
\providecommand \bibnamefont  [1]{#1}%
\providecommand \bibfnamefont [1]{#1}%
\providecommand \citenamefont [1]{#1}%
\providecommand \href@noop [0]{\@secondoftwo}%
\providecommand \href [0]{\begingroup \@sanitize@url \@href}%
\providecommand \@href[1]{\@@startlink{#1}\@@href}%
\providecommand \@@href[1]{\endgroup#1\@@endlink}%
\providecommand \@sanitize@url [0]{\catcode `\\12\catcode `\$12\catcode
  `\&12\catcode `\#12\catcode `\^12\catcode `\_12\catcode `\%12\relax}%
\providecommand \@@startlink[1]{}%
\providecommand \@@endlink[0]{}%
\providecommand \url  [0]{\begingroup\@sanitize@url \@url }%
\providecommand \@url [1]{\endgroup\@href {#1}{\urlprefix }}%
\providecommand \urlprefix  [0]{URL }%
\providecommand \Eprint [0]{\href }%
\providecommand \doibase [0]{https://doi.org/}%
\providecommand \selectlanguage [0]{\@gobble}%
\providecommand \bibinfo  [0]{\@secondoftwo}%
\providecommand \bibfield  [0]{\@secondoftwo}%
\providecommand \translation [1]{[#1]}%
\providecommand \BibitemOpen [0]{}%
\providecommand \bibitemStop [0]{}%
\providecommand \bibitemNoStop [0]{.\EOS\space}%
\providecommand \EOS [0]{\spacefactor3000\relax}%
\providecommand \BibitemShut  [1]{\csname bibitem#1\endcsname}%
\let\auto@bib@innerbib\@empty
\bibitem [{\citenamefont {Ashcroft}\ \emph {et~al.}(1976)\citenamefont
  {Ashcroft}, \citenamefont {Mermin} \emph {et~al.}}]{AshcroftBook}%
  \BibitemOpen
  \bibfield  {author} {\bibinfo {author} {\bibfnamefont {N.~W.}\ \bibnamefont
  {Ashcroft}}, \bibinfo {author} {\bibfnamefont {N.~D.}\ \bibnamefont
  {Mermin}}, \emph {et~al.},\ }\href@noop {} {\bibinfo {title} {Solid state
  physics}} (\bibinfo {year} {1976})\BibitemShut {NoStop}%
\bibitem [{\citenamefont {Stenger}\ \emph {et~al.}(1999)\citenamefont
  {Stenger}, \citenamefont {Inouye}, \citenamefont {Chikkatur}, \citenamefont
  {Stamper-Kurn}, \citenamefont {Pritchard},\ and\ \citenamefont
  {Ketterle}}]{Stenger_1999}%
  \BibitemOpen
  \bibfield  {author} {\bibinfo {author} {\bibfnamefont {J.}~\bibnamefont
  {Stenger}}, \bibinfo {author} {\bibfnamefont {S.}~\bibnamefont {Inouye}},
  \bibinfo {author} {\bibfnamefont {A.~P.}\ \bibnamefont {Chikkatur}}, \bibinfo
  {author} {\bibfnamefont {D.~M.}\ \bibnamefont {Stamper-Kurn}}, \bibinfo
  {author} {\bibfnamefont {D.~E.}\ \bibnamefont {Pritchard}},\ and\ \bibinfo
  {author} {\bibfnamefont {W.}~\bibnamefont {Ketterle}},\ }\bibfield  {title}
  {\bibinfo {title} {Bragg spectroscopy of a bose-einstein condensate},\ }\href
  {https://doi.org/10.1103/PhysRevLett.82.4569} {\bibfield  {journal} {\bibinfo
   {journal} {Phys. Rev. Lett.}\ }\textbf {\bibinfo {volume} {82}},\ \bibinfo
  {pages} {4569} (\bibinfo {year} {1999})}\BibitemShut {NoStop}%
\bibitem [{\citenamefont {Zoubi}\ and\ \citenamefont
  {Ritsch}(2010)}]{Zoubi_2010}%
  \BibitemOpen
  \bibfield  {author} {\bibinfo {author} {\bibfnamefont {H.}~\bibnamefont
  {Zoubi}}\ and\ \bibinfo {author} {\bibfnamefont {H.}~\bibnamefont {Ritsch}},\
  }\bibfield  {title} {\bibinfo {title} {Metastability and directional emission
  characteristics of excitons in 1d optical lattices},\ }\href
  {https://doi.org/10.1209/0295-5075/90/23001} {\bibfield  {journal} {\bibinfo
  {journal} {EPL}\ }\textbf {\bibinfo {volume} {90}},\ \bibinfo {pages} {23001}
  (\bibinfo {year} {2010})}\BibitemShut {NoStop}%
\bibitem [{\citenamefont {Weitenberg}\ \emph {et~al.}(2011)\citenamefont
  {Weitenberg}, \citenamefont {Schau\ss{}}, \citenamefont {Fukuhara},
  \citenamefont {Cheneau}, \citenamefont {Endres}, \citenamefont {Bloch},\ and\
  \citenamefont {Kuhr}}]{Weitenberg_2011}%
  \BibitemOpen
  \bibfield  {author} {\bibinfo {author} {\bibfnamefont {C.}~\bibnamefont
  {Weitenberg}}, \bibinfo {author} {\bibfnamefont {P.}~\bibnamefont
  {Schau\ss{}}}, \bibinfo {author} {\bibfnamefont {T.}~\bibnamefont
  {Fukuhara}}, \bibinfo {author} {\bibfnamefont {M.}~\bibnamefont {Cheneau}},
  \bibinfo {author} {\bibfnamefont {M.}~\bibnamefont {Endres}}, \bibinfo
  {author} {\bibfnamefont {I.}~\bibnamefont {Bloch}},\ and\ \bibinfo {author}
  {\bibfnamefont {S.}~\bibnamefont {Kuhr}},\ }\bibfield  {title} {\bibinfo
  {title} {Coherent light scattering from a two-dimensional mott insulator},\
  }\href {https://doi.org/10.1103/PhysRevLett.106.215301} {\bibfield  {journal}
  {\bibinfo  {journal} {Phys. Rev. Lett.}\ }\textbf {\bibinfo {volume} {106}},\
  \bibinfo {pages} {215301} (\bibinfo {year} {2011})}\BibitemShut {NoStop}%
\bibitem [{\citenamefont {Chang}\ \emph {et~al.}(2012)\citenamefont {Chang},
  \citenamefont {Jiang}, \citenamefont {Gorshkov},\ and\ \citenamefont
  {Kimble}}]{Chang_2012}%
  \BibitemOpen
  \bibfield  {author} {\bibinfo {author} {\bibfnamefont {D.~E.}\ \bibnamefont
  {Chang}}, \bibinfo {author} {\bibfnamefont {L.}~\bibnamefont {Jiang}},
  \bibinfo {author} {\bibfnamefont {A.~V.}\ \bibnamefont {Gorshkov}},\ and\
  \bibinfo {author} {\bibfnamefont {H.~J.}\ \bibnamefont {Kimble}},\ }\bibfield
   {title} {\bibinfo {title} {Cavity {QED} with atomic mirrors},\ }\href
  {https://doi.org/10.1088/1367-2630/14/6/063003} {\bibfield  {journal}
  {\bibinfo  {journal} {New J. Phys.}\ }\textbf {\bibinfo {volume} {14}},\
  \bibinfo {pages} {063003} (\bibinfo {year} {2012})}\BibitemShut {NoStop}%
\bibitem [{\citenamefont {Kornovan}\ \emph {et~al.}(2016)\citenamefont
  {Kornovan}, \citenamefont {Sheremet},\ and\ \citenamefont
  {Petrov}}]{Kornovan_2016}%
  \BibitemOpen
  \bibfield  {author} {\bibinfo {author} {\bibfnamefont {D.~F.}\ \bibnamefont
  {Kornovan}}, \bibinfo {author} {\bibfnamefont {A.~S.}\ \bibnamefont
  {Sheremet}},\ and\ \bibinfo {author} {\bibfnamefont {M.~I.}\ \bibnamefont
  {Petrov}},\ }\bibfield  {title} {\bibinfo {title} {Collective polaritonic
  modes in an array of two-level quantum emitters coupled to an optical
  nanofiber},\ }\href {https://doi.org/10.1103/PhysRevB.94.245416} {\bibfield
  {journal} {\bibinfo  {journal} {Phys. Rev. B}\ }\textbf {\bibinfo {volume}
  {94}},\ \bibinfo {pages} {245416} (\bibinfo {year} {2016})}\BibitemShut
  {NoStop}%
\bibitem [{\citenamefont {Corzo}\ \emph {et~al.}(2016)\citenamefont {Corzo},
  \citenamefont {Gouraud}, \citenamefont {Chandra}, \citenamefont {Goban},
  \citenamefont {Sheremet}, \citenamefont {Kupriyanov},\ and\ \citenamefont
  {Laurat}}]{Corzo_2016}%
  \BibitemOpen
  \bibfield  {author} {\bibinfo {author} {\bibfnamefont {N.~V.}\ \bibnamefont
  {Corzo}}, \bibinfo {author} {\bibfnamefont {B.}~\bibnamefont {Gouraud}},
  \bibinfo {author} {\bibfnamefont {A.}~\bibnamefont {Chandra}}, \bibinfo
  {author} {\bibfnamefont {A.}~\bibnamefont {Goban}}, \bibinfo {author}
  {\bibfnamefont {A.~S.}\ \bibnamefont {Sheremet}}, \bibinfo {author}
  {\bibfnamefont {D.~V.}\ \bibnamefont {Kupriyanov}},\ and\ \bibinfo {author}
  {\bibfnamefont {J.}~\bibnamefont {Laurat}},\ }\bibfield  {title} {\bibinfo
  {title} {Large bragg reflection from one-dimensional chains of trapped atoms
  near a nanoscale waveguide},\ }\href
  {https://doi.org/10.1103/PhysRevLett.117.133603} {\bibfield  {journal}
  {\bibinfo  {journal} {Phys. Rev. Lett.}\ }\textbf {\bibinfo {volume} {117}},\
  \bibinfo {pages} {133603} (\bibinfo {year} {2016})}\BibitemShut {NoStop}%
\bibitem [{\citenamefont {S\o{}rensen}\ \emph {et~al.}(2016)\citenamefont
  {S\o{}rensen}, \citenamefont {B\'eguin}, \citenamefont {Kluge}, \citenamefont
  {Iakoupov}, \citenamefont {S\o{}rensen}, \citenamefont {M\"uller},
  \citenamefont {Polzik},\ and\ \citenamefont {Appel}}]{Soerensen_2016}%
  \BibitemOpen
  \bibfield  {author} {\bibinfo {author} {\bibfnamefont {H.~L.}\ \bibnamefont
  {S\o{}rensen}}, \bibinfo {author} {\bibfnamefont {J.-B.}\ \bibnamefont
  {B\'eguin}}, \bibinfo {author} {\bibfnamefont {K.~W.}\ \bibnamefont {Kluge}},
  \bibinfo {author} {\bibfnamefont {I.}~\bibnamefont {Iakoupov}}, \bibinfo
  {author} {\bibfnamefont {A.~S.}\ \bibnamefont {S\o{}rensen}}, \bibinfo
  {author} {\bibfnamefont {J.~H.}\ \bibnamefont {M\"uller}}, \bibinfo {author}
  {\bibfnamefont {E.~S.}\ \bibnamefont {Polzik}},\ and\ \bibinfo {author}
  {\bibfnamefont {J.}~\bibnamefont {Appel}},\ }\bibfield  {title} {\bibinfo
  {title} {Coherent backscattering of light off one-dimensional atomic
  strings},\ }\href {https://doi.org/10.1103/PhysRevLett.117.133604} {\bibfield
   {journal} {\bibinfo  {journal} {Phys. Rev. Lett.}\ }\textbf {\bibinfo
  {volume} {117}},\ \bibinfo {pages} {133604} (\bibinfo {year}
  {2016})}\BibitemShut {NoStop}%
\bibitem [{\citenamefont {Olmos}\ \emph {et~al.}(2020)\citenamefont {Olmos},
  \citenamefont {Buonaiuto}, \citenamefont {Schneeweiss},\ and\ \citenamefont
  {Lesanovsky}}]{Olmos_2020}%
  \BibitemOpen
  \bibfield  {author} {\bibinfo {author} {\bibfnamefont {B.}~\bibnamefont
  {Olmos}}, \bibinfo {author} {\bibfnamefont {G.}~\bibnamefont {Buonaiuto}},
  \bibinfo {author} {\bibfnamefont {P.}~\bibnamefont {Schneeweiss}},\ and\
  \bibinfo {author} {\bibfnamefont {I.}~\bibnamefont {Lesanovsky}},\ }\bibfield
   {title} {\bibinfo {title} {Interaction signatures and non-gaussian photon
  states from a strongly driven atomic ensemble coupled to a nanophotonic
  waveguide},\ }\href {https://doi.org/10.1103/PhysRevA.102.043711} {\bibfield
  {journal} {\bibinfo  {journal} {Phys. Rev. A}\ }\textbf {\bibinfo {volume}
  {102}},\ \bibinfo {pages} {043711} (\bibinfo {year} {2020})}\BibitemShut
  {NoStop}%
\bibitem [{\citenamefont {Meng}\ \emph {et~al.}(2020)\citenamefont {Meng},
  \citenamefont {Liedl}, \citenamefont {Pucher}, \citenamefont
  {Rauschenbeutel},\ and\ \citenamefont {Schneeweiss}}]{Meng_2020}%
  \BibitemOpen
  \bibfield  {author} {\bibinfo {author} {\bibfnamefont {Y.}~\bibnamefont
  {Meng}}, \bibinfo {author} {\bibfnamefont {C.}~\bibnamefont {Liedl}},
  \bibinfo {author} {\bibfnamefont {S.}~\bibnamefont {Pucher}}, \bibinfo
  {author} {\bibfnamefont {A.}~\bibnamefont {Rauschenbeutel}},\ and\ \bibinfo
  {author} {\bibfnamefont {P.}~\bibnamefont {Schneeweiss}},\ }\bibfield
  {title} {\bibinfo {title} {Imaging and localizing individual atoms interfaced
  with a nanophotonic waveguide},\ }\href
  {https://doi.org/10.1103/PhysRevLett.125.053603} {\bibfield  {journal}
  {\bibinfo  {journal} {Phys. Rev. Lett.}\ }\textbf {\bibinfo {volume} {125}},\
  \bibinfo {pages} {053603} (\bibinfo {year} {2020})}\BibitemShut {NoStop}%
\bibitem [{\citenamefont {Kaiser}(2009)}]{Kaiser_2009}%
  \BibitemOpen
  \bibfield  {author} {\bibinfo {author} {\bibfnamefont {R.}~\bibnamefont
  {Kaiser}},\ }\bibfield  {title} {\bibinfo {title} {Quantum multiple
  scattering},\ }\href {https://doi.org/10.1080/09500340903082663} {\bibfield
  {journal} {\bibinfo  {journal} {J. Mod. Opt.}\ }\textbf {\bibinfo {volume}
  {56}},\ \bibinfo {pages} {2082–2088} (\bibinfo {year} {2009})}\BibitemShut
  {NoStop}%
\bibitem [{\citenamefont {Sokolov}\ and\ \citenamefont
  {Guerin}(2019)}]{Sokolov_2019}%
  \BibitemOpen
  \bibfield  {author} {\bibinfo {author} {\bibfnamefont {I.~M.}\ \bibnamefont
  {Sokolov}}\ and\ \bibinfo {author} {\bibfnamefont {W.}~\bibnamefont
  {Guerin}},\ }\bibfield  {title} {\bibinfo {title} {Comparison of three
  approaches to light scattering by dilute cold atomic ensembles},\ }\href@noop
  {} {\bibfield  {journal} {\bibinfo  {journal} {JOSA B}\ }\textbf {\bibinfo
  {volume} {36}},\ \bibinfo {pages} {2030} (\bibinfo {year}
  {2019})}\BibitemShut {NoStop}%
\bibitem [{\citenamefont {Kolchin}\ \emph {et~al.}(2011)\citenamefont
  {Kolchin}, \citenamefont {Oulton},\ and\ \citenamefont
  {Zhang}}]{Kolchin_2011}%
  \BibitemOpen
  \bibfield  {author} {\bibinfo {author} {\bibfnamefont {P.}~\bibnamefont
  {Kolchin}}, \bibinfo {author} {\bibfnamefont {R.~F.}\ \bibnamefont
  {Oulton}},\ and\ \bibinfo {author} {\bibfnamefont {X.}~\bibnamefont
  {Zhang}},\ }\bibfield  {title} {\bibinfo {title} {Nonlinear quantum optics in
  a waveguide: Distinct single photons strongly interacting at the single atom
  level},\ }\href {https://doi.org/10.1103/PhysRevLett.106.113601} {\bibfield
  {journal} {\bibinfo  {journal} {Phys. Rev. Lett.}\ }\textbf {\bibinfo
  {volume} {106}},\ \bibinfo {pages} {113601} (\bibinfo {year}
  {2011})}\BibitemShut {NoStop}%
\bibitem [{\citenamefont {Chang}\ \emph {et~al.}(2014)\citenamefont {Chang},
  \citenamefont {Vuletic},\ and\ \citenamefont {Lukin}}]{Chang_2014}%
  \BibitemOpen
  \bibfield  {author} {\bibinfo {author} {\bibfnamefont {D.~E.}\ \bibnamefont
  {Chang}}, \bibinfo {author} {\bibfnamefont {V.}~\bibnamefont {Vuletic}},\
  and\ \bibinfo {author} {\bibfnamefont {M.~D.}\ \bibnamefont {Lukin}},\
  }\bibfield  {title} {\bibinfo {title} {Quantum nonlinear optics — photon by
  photon},\ }\href {https://doi.org/10.1038/nphoton.2014.192} {\bibfield
  {journal} {\bibinfo  {journal} {Nature Photon.}\ }\textbf {\bibinfo {volume}
  {8}},\ \bibinfo {pages} {685} (\bibinfo {year} {2014})}\BibitemShut {NoStop}%
\bibitem [{\citenamefont {Firstenberg}\ \emph {et~al.}(2016)\citenamefont
  {Firstenberg}, \citenamefont {Adams},\ and\ \citenamefont
  {Hofferberth}}]{Firstenberg_2016}%
  \BibitemOpen
  \bibfield  {author} {\bibinfo {author} {\bibfnamefont {O.}~\bibnamefont
  {Firstenberg}}, \bibinfo {author} {\bibfnamefont {C.~S.}\ \bibnamefont
  {Adams}},\ and\ \bibinfo {author} {\bibfnamefont {S.}~\bibnamefont
  {Hofferberth}},\ }\bibfield  {title} {\bibinfo {title} {Nonlinear quantum
  optics mediated by rydberg interactions},\ }\href
  {https://doi.org/10.1088/0953-4075/49/15/152003} {\bibfield  {journal}
  {\bibinfo  {journal} {J. Phys. B: At. Mol. Opt. Phys.}\ }\textbf {\bibinfo
  {volume} {49}},\ \bibinfo {pages} {152003} (\bibinfo {year}
  {2016})}\BibitemShut {NoStop}%
\bibitem [{\citenamefont {Le~Kien}\ and\ \citenamefont
  {Rauschenbeutel}(2014)}]{LeKien_2014}%
  \BibitemOpen
  \bibfield  {author} {\bibinfo {author} {\bibfnamefont {F.}~\bibnamefont
  {Le~Kien}}\ and\ \bibinfo {author} {\bibfnamefont {A.}~\bibnamefont
  {Rauschenbeutel}},\ }\bibfield  {title} {\bibinfo {title} {Propagation of
  nanofiber-guided light through an array of atoms},\ }\href
  {https://doi.org/10.1103/PhysRevA.90.063816} {\bibfield  {journal} {\bibinfo
  {journal} {Phys. Rev. A}\ }\textbf {\bibinfo {volume} {90}},\ \bibinfo
  {pages} {063816} (\bibinfo {year} {2014})}\BibitemShut {NoStop}%
\bibitem [{\citenamefont {Bettles}\ \emph {et~al.}(2016)\citenamefont
  {Bettles}, \citenamefont {Gardiner},\ and\ \citenamefont
  {Adams}}]{Bettles_2016}%
  \BibitemOpen
  \bibfield  {author} {\bibinfo {author} {\bibfnamefont {R.~J.}\ \bibnamefont
  {Bettles}}, \bibinfo {author} {\bibfnamefont {S.~A.}\ \bibnamefont
  {Gardiner}},\ and\ \bibinfo {author} {\bibfnamefont {C.~S.}\ \bibnamefont
  {Adams}},\ }\bibfield  {title} {\bibinfo {title} {Enhanced optical cross
  section via collective coupling of atomic dipoles in a 2d array},\ }\href
  {https://doi.org/10.1103/PhysRevLett.116.103602} {\bibfield  {journal}
  {\bibinfo  {journal} {Phys. Rev. Lett.}\ }\textbf {\bibinfo {volume} {116}},\
  \bibinfo {pages} {103602} (\bibinfo {year} {2016})}\BibitemShut {NoStop}%
\bibitem [{\citenamefont {Rui}\ \emph {et~al.}(2020)\citenamefont {Rui},
  \citenamefont {Wei}, \citenamefont {Rubio-Abadal}, \citenamefont {Hollerith},
  \citenamefont {Zeiher}, \citenamefont {Stamper-Kurn}, \citenamefont {Gross},\
  and\ \citenamefont {Bloch}}]{Rui_2020}%
  \BibitemOpen
  \bibfield  {author} {\bibinfo {author} {\bibfnamefont {J.}~\bibnamefont
  {Rui}}, \bibinfo {author} {\bibfnamefont {D.}~\bibnamefont {Wei}}, \bibinfo
  {author} {\bibfnamefont {A.}~\bibnamefont {Rubio-Abadal}}, \bibinfo {author}
  {\bibfnamefont {S.}~\bibnamefont {Hollerith}}, \bibinfo {author}
  {\bibfnamefont {J.}~\bibnamefont {Zeiher}}, \bibinfo {author} {\bibfnamefont
  {D.~M.}\ \bibnamefont {Stamper-Kurn}}, \bibinfo {author} {\bibfnamefont
  {C.}~\bibnamefont {Gross}},\ and\ \bibinfo {author} {\bibfnamefont
  {I.}~\bibnamefont {Bloch}},\ }\bibfield  {title} {\bibinfo {title} {A
  subradiant optical mirror formed by a single structured atomic layer},\
  }\href {https://doi.org/10.1038/s41586-020-2463-x} {\bibfield  {journal}
  {\bibinfo  {journal} {Nature}\ }\textbf {\bibinfo {volume} {583}},\ \bibinfo
  {pages} {369–374} (\bibinfo {year} {2020})}\BibitemShut {NoStop}%
\bibitem [{\citenamefont {Facchinetti}\ \emph {et~al.}(2016)\citenamefont
  {Facchinetti}, \citenamefont {Jenkins},\ and\ \citenamefont
  {Ruostekoski}}]{Facchinetti_2016}%
  \BibitemOpen
  \bibfield  {author} {\bibinfo {author} {\bibfnamefont {G.}~\bibnamefont
  {Facchinetti}}, \bibinfo {author} {\bibfnamefont {S.~D.}\ \bibnamefont
  {Jenkins}},\ and\ \bibinfo {author} {\bibfnamefont {J.}~\bibnamefont
  {Ruostekoski}},\ }\bibfield  {title} {\bibinfo {title} {Storing light with
  subradiant correlations in arrays of atoms},\ }\href
  {https://doi.org/10.1103/PhysRevLett.117.243601} {\bibfield  {journal}
  {\bibinfo  {journal} {Phys. Rev. Lett.}\ }\textbf {\bibinfo {volume} {117}},\
  \bibinfo {pages} {243601} (\bibinfo {year} {2016})}\BibitemShut {NoStop}%
\bibitem [{\citenamefont {Asenjo-Garcia}\ \emph
  {et~al.}(2017{\natexlab{a}})\citenamefont {Asenjo-Garcia}, \citenamefont
  {Moreno-Cardoner}, \citenamefont {Albrecht}, \citenamefont {Kimble},\ and\
  \citenamefont {Chang}}]{Asenjo_2017}%
  \BibitemOpen
  \bibfield  {author} {\bibinfo {author} {\bibfnamefont {A.}~\bibnamefont
  {Asenjo-Garcia}}, \bibinfo {author} {\bibfnamefont {M.}~\bibnamefont
  {Moreno-Cardoner}}, \bibinfo {author} {\bibfnamefont {A.}~\bibnamefont
  {Albrecht}}, \bibinfo {author} {\bibfnamefont {H.~J.}\ \bibnamefont
  {Kimble}},\ and\ \bibinfo {author} {\bibfnamefont {D.~E.}\ \bibnamefont
  {Chang}},\ }\bibfield  {title} {\bibinfo {title} {Exponential improvement in
  photon storage fidelities using subradiance and ``selective radiance'' in
  atomic arrays},\ }\href {https://doi.org/10.1103/PhysRevX.7.031024}
  {\bibfield  {journal} {\bibinfo  {journal} {Phys. Rev. X}\ }\textbf {\bibinfo
  {volume} {7}},\ \bibinfo {pages} {031024} (\bibinfo {year}
  {2017}{\natexlab{a}})}\BibitemShut {NoStop}%
\bibitem [{\citenamefont {Asenjo-Garcia}\ \emph {et~al.}(2019)\citenamefont
  {Asenjo-Garcia}, \citenamefont {Kimble},\ and\ \citenamefont
  {Chang}}]{Asenjo_2019}%
  \BibitemOpen
  \bibfield  {author} {\bibinfo {author} {\bibfnamefont {A.}~\bibnamefont
  {Asenjo-Garcia}}, \bibinfo {author} {\bibfnamefont {H.~J.}\ \bibnamefont
  {Kimble}},\ and\ \bibinfo {author} {\bibfnamefont {D.~E.}\ \bibnamefont
  {Chang}},\ }\bibfield  {title} {\bibinfo {title} {Optical waveguiding by
  atomic entanglement in multilevel atom arrays},\ }\href
  {https://doi.org/10.1073/pnas.1911467116} {\bibfield  {journal} {\bibinfo
  {journal} {PNAS}\ }\textbf {\bibinfo {volume} {116}},\ \bibinfo {pages}
  {25503} (\bibinfo {year} {2019})}\BibitemShut {NoStop}%
\bibitem [{\citenamefont {Masson}\ and\ \citenamefont
  {Asenjo-Garcia}(2020)}]{Masson_2020}%
  \BibitemOpen
  \bibfield  {author} {\bibinfo {author} {\bibfnamefont {S.~J.}\ \bibnamefont
  {Masson}}\ and\ \bibinfo {author} {\bibfnamefont {A.}~\bibnamefont
  {Asenjo-Garcia}},\ }\bibfield  {title} {\bibinfo {title} {Atomic-waveguide
  quantum electrodynamics},\ }\href
  {https://doi.org/10.1103/PhysRevResearch.2.043213} {\bibfield  {journal}
  {\bibinfo  {journal} {Phys. Rev. Research}\ }\textbf {\bibinfo {volume}
  {2}},\ \bibinfo {pages} {043213} (\bibinfo {year} {2020})}\BibitemShut
  {NoStop}%
\bibitem [{\citenamefont {Jones}\ \emph {et~al.}(2020)\citenamefont {Jones},
  \citenamefont {Buonaiuto}, \citenamefont {Lang}, \citenamefont {Lesanovsky},\
  and\ \citenamefont {Olmos}}]{Jones_2020}%
  \BibitemOpen
  \bibfield  {author} {\bibinfo {author} {\bibfnamefont {R.}~\bibnamefont
  {Jones}}, \bibinfo {author} {\bibfnamefont {G.}~\bibnamefont {Buonaiuto}},
  \bibinfo {author} {\bibfnamefont {B.}~\bibnamefont {Lang}}, \bibinfo {author}
  {\bibfnamefont {I.}~\bibnamefont {Lesanovsky}},\ and\ \bibinfo {author}
  {\bibfnamefont {B.}~\bibnamefont {Olmos}},\ }\bibfield  {title} {\bibinfo
  {title} {Collectively enhanced chiral photon emission from an atomic array
  near a nanofiber},\ }\href {https://doi.org/10.1103/PhysRevLett.124.093601}
  {\bibfield  {journal} {\bibinfo  {journal} {Phys. Rev. Lett.}\ }\textbf
  {\bibinfo {volume} {124}},\ \bibinfo {pages} {093601} (\bibinfo {year}
  {2020})}\BibitemShut {NoStop}%
\bibitem [{\citenamefont {Birkl}\ \emph {et~al.}(1995)\citenamefont {Birkl},
  \citenamefont {Gatzke}, \citenamefont {Deutsch}, \citenamefont {Rolston},\
  and\ \citenamefont {Phillips}}]{Birkl_1995}%
  \BibitemOpen
  \bibfield  {author} {\bibinfo {author} {\bibfnamefont {G.}~\bibnamefont
  {Birkl}}, \bibinfo {author} {\bibfnamefont {M.}~\bibnamefont {Gatzke}},
  \bibinfo {author} {\bibfnamefont {I.~H.}\ \bibnamefont {Deutsch}}, \bibinfo
  {author} {\bibfnamefont {S.~L.}\ \bibnamefont {Rolston}},\ and\ \bibinfo
  {author} {\bibfnamefont {W.~D.}\ \bibnamefont {Phillips}},\ }\bibfield
  {title} {\bibinfo {title} {Bragg scattering from atoms in optical lattices},\
  }\href {https://doi.org/10.1103/PhysRevLett.75.2823} {\bibfield  {journal}
  {\bibinfo  {journal} {Phys. Rev. Lett.}\ }\textbf {\bibinfo {volume} {75}},\
  \bibinfo {pages} {2823} (\bibinfo {year} {1995})}\BibitemShut {NoStop}%
\bibitem [{\citenamefont {Lodahl}\ \emph {et~al.}(2017)\citenamefont {Lodahl},
  \citenamefont {Mahmoodian}, \citenamefont {Stobbe}, \citenamefont
  {Rauschenbeutel}, \citenamefont {Schneeweiss}, \citenamefont {Volz},
  \citenamefont {Pichler},\ and\ \citenamefont {Zoller}}]{Lodahl_2017}%
  \BibitemOpen
  \bibfield  {author} {\bibinfo {author} {\bibfnamefont {P.}~\bibnamefont
  {Lodahl}}, \bibinfo {author} {\bibfnamefont {S.}~\bibnamefont {Mahmoodian}},
  \bibinfo {author} {\bibfnamefont {S.}~\bibnamefont {Stobbe}}, \bibinfo
  {author} {\bibfnamefont {A.}~\bibnamefont {Rauschenbeutel}}, \bibinfo
  {author} {\bibfnamefont {P.}~\bibnamefont {Schneeweiss}}, \bibinfo {author}
  {\bibfnamefont {J.}~\bibnamefont {Volz}}, \bibinfo {author} {\bibfnamefont
  {H.}~\bibnamefont {Pichler}},\ and\ \bibinfo {author} {\bibfnamefont
  {P.}~\bibnamefont {Zoller}},\ }\bibfield  {title} {\bibinfo {title} {Chiral
  quantum optics},\ }\href {https://doi.org/10.1038/nature21037} {\bibfield
  {journal} {\bibinfo  {journal} {Nature}\ }\textbf {\bibinfo {volume} {541}},\
  \bibinfo {pages} {473} (\bibinfo {year} {2017})}\BibitemShut {NoStop}%
\bibitem [{\citenamefont {Le~Kien}\ and\ \citenamefont
  {Rauschenbeutel}(2017)}]{LeKien_2017}%
  \BibitemOpen
  \bibfield  {author} {\bibinfo {author} {\bibfnamefont {F.}~\bibnamefont
  {Le~Kien}}\ and\ \bibinfo {author} {\bibfnamefont {A.}~\bibnamefont
  {Rauschenbeutel}},\ }\bibfield  {title} {\bibinfo {title} {Nanofiber-mediated
  chiral radiative coupling between two atoms},\ }\href
  {https://doi.org/10.1103/PhysRevA.95.023838} {\bibfield  {journal} {\bibinfo
  {journal} {Phys. Rev. A}\ }\textbf {\bibinfo {volume} {95}},\ \bibinfo
  {pages} {023838} (\bibinfo {year} {2017})}\BibitemShut {NoStop}%
\bibitem [{\citenamefont {Asenjo-Garcia}\ \emph
  {et~al.}(2017{\natexlab{b}})\citenamefont {Asenjo-Garcia}, \citenamefont
  {Hood}, \citenamefont {Chang},\ and\ \citenamefont {Kimble}}]{Asenjo_2017b}%
  \BibitemOpen
  \bibfield  {author} {\bibinfo {author} {\bibfnamefont {A.}~\bibnamefont
  {Asenjo-Garcia}}, \bibinfo {author} {\bibfnamefont {J.~D.}\ \bibnamefont
  {Hood}}, \bibinfo {author} {\bibfnamefont {D.~E.}\ \bibnamefont {Chang}},\
  and\ \bibinfo {author} {\bibfnamefont {H.~J.}\ \bibnamefont {Kimble}},\
  }\bibfield  {title} {\bibinfo {title} {Atom-light interactions in
  quasi-one-dimensional nanostructures: A green's-function perspective},\
  }\href {https://doi.org/10.1103/PhysRevA.95.033818} {\bibfield  {journal}
  {\bibinfo  {journal} {Phys. Rev. A}\ }\textbf {\bibinfo {volume} {95}},\
  \bibinfo {pages} {033818} (\bibinfo {year} {2017}{\natexlab{b}})}\BibitemShut
  {NoStop}%
\bibitem [{\citenamefont {Le~Kien}\ \emph {et~al.}(2005)\citenamefont
  {Le~Kien}, \citenamefont {Dutta~Gupta}, \citenamefont {Balykin},\ and\
  \citenamefont {Hakuta}}]{LeKien_2005}%
  \BibitemOpen
  \bibfield  {author} {\bibinfo {author} {\bibfnamefont {F.}~\bibnamefont
  {Le~Kien}}, \bibinfo {author} {\bibfnamefont {S.}~\bibnamefont
  {Dutta~Gupta}}, \bibinfo {author} {\bibfnamefont {V.~I.}\ \bibnamefont
  {Balykin}},\ and\ \bibinfo {author} {\bibfnamefont {K.}~\bibnamefont
  {Hakuta}},\ }\bibfield  {title} {\bibinfo {title} {Spontaneous emission of a
  cesium atom near a nanofiber: Efficient coupling of light to guided modes},\
  }\href {https://doi.org/10.1103/PhysRevA.72.032509} {\bibfield  {journal}
  {\bibinfo  {journal} {Phys. Rev. A}\ }\textbf {\bibinfo {volume} {72}},\
  \bibinfo {pages} {032509} (\bibinfo {year} {2005})}\BibitemShut {NoStop}%
\bibitem [{\citenamefont {Scheel}\ \emph {et~al.}(2015)\citenamefont {Scheel},
  \citenamefont {Buhmann}, \citenamefont {Clausen},\ and\ \citenamefont
  {Schneeweiss}}]{Scheel_2015}%
  \BibitemOpen
  \bibfield  {author} {\bibinfo {author} {\bibfnamefont {S.}~\bibnamefont
  {Scheel}}, \bibinfo {author} {\bibfnamefont {S.~Y.}\ \bibnamefont {Buhmann}},
  \bibinfo {author} {\bibfnamefont {C.}~\bibnamefont {Clausen}},\ and\ \bibinfo
  {author} {\bibfnamefont {P.}~\bibnamefont {Schneeweiss}},\ }\bibfield
  {title} {\bibinfo {title} {Directional spontaneous emission and lateral
  casimir-polder force on an atom close to a nanofiber},\ }\href
  {https://doi.org/10.1103/PhysRevA.92.043819} {\bibfield  {journal} {\bibinfo
  {journal} {Phys. Rev. A}\ }\textbf {\bibinfo {volume} {92}},\ \bibinfo
  {pages} {043819} (\bibinfo {year} {2015})}\BibitemShut {NoStop}%
\bibitem [{\citenamefont {Solano}\ \emph {et~al.}(2017)\citenamefont {Solano},
  \citenamefont {Barberis-Blostein}, \citenamefont {Fatemi}, \citenamefont
  {Orozco},\ and\ \citenamefont {Rolston}}]{Solano_2017b}%
  \BibitemOpen
  \bibfield  {author} {\bibinfo {author} {\bibfnamefont {P.}~\bibnamefont
  {Solano}}, \bibinfo {author} {\bibfnamefont {P.}~\bibnamefont
  {Barberis-Blostein}}, \bibinfo {author} {\bibfnamefont {F.~K.}\ \bibnamefont
  {Fatemi}}, \bibinfo {author} {\bibfnamefont {L.~A.}\ \bibnamefont {Orozco}},\
  and\ \bibinfo {author} {\bibfnamefont {S.~L.}\ \bibnamefont {Rolston}},\
  }\bibfield  {title} {\bibinfo {title} {Super-radiance reveals infinite-range
  dipole interactions through a nanofiber},\ }\href@noop {} {\bibfield
  {journal} {\bibinfo  {journal} {Nature communications}\ }\textbf {\bibinfo
  {volume} {8}},\ \bibinfo {pages} {1857} (\bibinfo {year} {2017})}\BibitemShut
  {NoStop}%
\bibitem [{\citenamefont {Zhang}\ \emph {et~al.}(2020)\citenamefont {Zhang},
  \citenamefont {Yu},\ and\ \citenamefont {M\o{}lmer}}]{Zhang_2020}%
  \BibitemOpen
  \bibfield  {author} {\bibinfo {author} {\bibfnamefont {Y.-X.}\ \bibnamefont
  {Zhang}}, \bibinfo {author} {\bibfnamefont {C.}~\bibnamefont {Yu}},\ and\
  \bibinfo {author} {\bibfnamefont {K.}~\bibnamefont {M\o{}lmer}},\ }\bibfield
  {title} {\bibinfo {title} {Subradiant bound dimer excited states of emitter
  chains coupled to a one dimensional waveguide},\ }\href
  {https://doi.org/10.1103/PhysRevResearch.2.013173} {\bibfield  {journal}
  {\bibinfo  {journal} {Phys. Rev. Research}\ }\textbf {\bibinfo {volume}
  {2}},\ \bibinfo {pages} {013173} (\bibinfo {year} {2020})}\BibitemShut
  {NoStop}%
\bibitem [{\citenamefont {Ramos}\ \emph {et~al.}(2014)\citenamefont {Ramos},
  \citenamefont {Pichler}, \citenamefont {Daley},\ and\ \citenamefont
  {Zoller}}]{Ramos_2014}%
  \BibitemOpen
  \bibfield  {author} {\bibinfo {author} {\bibfnamefont {T.}~\bibnamefont
  {Ramos}}, \bibinfo {author} {\bibfnamefont {H.}~\bibnamefont {Pichler}},
  \bibinfo {author} {\bibfnamefont {A.~J.}\ \bibnamefont {Daley}},\ and\
  \bibinfo {author} {\bibfnamefont {P.}~\bibnamefont {Zoller}},\ }\bibfield
  {title} {\bibinfo {title} {Quantum spin dimers from chiral dissipation in
  cold-atom chains},\ }\href {https://doi.org/10.1103/PhysRevLett.113.237203}
  {\bibfield  {journal} {\bibinfo  {journal} {Phys. Rev. Lett.}\ }\textbf
  {\bibinfo {volume} {113}},\ \bibinfo {pages} {237203} (\bibinfo {year}
  {2014})}\BibitemShut {NoStop}%
\bibitem [{\citenamefont {Pichler}\ \emph {et~al.}(2015)\citenamefont
  {Pichler}, \citenamefont {Ramos}, \citenamefont {Daley},\ and\ \citenamefont
  {Zoller}}]{Pichler_2015}%
  \BibitemOpen
  \bibfield  {author} {\bibinfo {author} {\bibfnamefont {H.}~\bibnamefont
  {Pichler}}, \bibinfo {author} {\bibfnamefont {T.}~\bibnamefont {Ramos}},
  \bibinfo {author} {\bibfnamefont {A.~J.}\ \bibnamefont {Daley}},\ and\
  \bibinfo {author} {\bibfnamefont {P.}~\bibnamefont {Zoller}},\ }\bibfield
  {title} {\bibinfo {title} {Quantum optics of chiral spin networks},\ }\href
  {https://doi.org/10.1103/PhysRevA.91.042116} {\bibfield  {journal} {\bibinfo
  {journal} {Phys. Rev. A}\ }\textbf {\bibinfo {volume} {91}},\ \bibinfo
  {pages} {042116} (\bibinfo {year} {2015})}\BibitemShut {NoStop}%
\bibitem [{\citenamefont {Vetsch}\ \emph {et~al.}(2010)\citenamefont {Vetsch},
  \citenamefont {Reitz}, \citenamefont {Sagu\'e}, \citenamefont {Schmidt},
  \citenamefont {Dawkins},\ and\ \citenamefont {Rauschenbeutel}}]{Vetsch_2010}%
  \BibitemOpen
  \bibfield  {author} {\bibinfo {author} {\bibfnamefont {E.}~\bibnamefont
  {Vetsch}}, \bibinfo {author} {\bibfnamefont {D.}~\bibnamefont {Reitz}},
  \bibinfo {author} {\bibfnamefont {G.}~\bibnamefont {Sagu\'e}}, \bibinfo
  {author} {\bibfnamefont {R.}~\bibnamefont {Schmidt}}, \bibinfo {author}
  {\bibfnamefont {S.~T.}\ \bibnamefont {Dawkins}},\ and\ \bibinfo {author}
  {\bibfnamefont {A.}~\bibnamefont {Rauschenbeutel}},\ }\bibfield  {title}
  {\bibinfo {title} {Optical interface created by laser-cooled atoms trapped in
  the evanescent field surrounding an optical nanofiber},\ }\href
  {https://doi.org/10.1103/PhysRevLett.104.203603} {\bibfield  {journal}
  {\bibinfo  {journal} {Phys. Rev. Lett.}\ }\textbf {\bibinfo {volume} {104}},\
  \bibinfo {pages} {203603} (\bibinfo {year} {2010})}\BibitemShut {NoStop}%
\bibitem [{\citenamefont {{Vetsch}}\ \emph {et~al.}(2012)\citenamefont
  {{Vetsch}}, \citenamefont {{Dawkins}}, \citenamefont {{Mitsch}},
  \citenamefont {{Reitz}}, \citenamefont {{Schneeweiss}},\ and\ \citenamefont
  {{Rauschenbeutel}}}]{Vetsch_2012}%
  \BibitemOpen
  \bibfield  {author} {\bibinfo {author} {\bibfnamefont {E.}~\bibnamefont
  {{Vetsch}}}, \bibinfo {author} {\bibfnamefont {S.~T.}\ \bibnamefont
  {{Dawkins}}}, \bibinfo {author} {\bibfnamefont {R.}~\bibnamefont {{Mitsch}}},
  \bibinfo {author} {\bibfnamefont {D.}~\bibnamefont {{Reitz}}}, \bibinfo
  {author} {\bibfnamefont {P.}~\bibnamefont {{Schneeweiss}}},\ and\ \bibinfo
  {author} {\bibfnamefont {A.}~\bibnamefont {{Rauschenbeutel}}},\ }\bibfield
  {title} {\bibinfo {title} {Nanofiber-based optical trapping of cold neutral
  atoms},\ }\href {https://doi.org/10.1109/JSTQE.2012.2196025} {\bibfield
  {journal} {\bibinfo  {journal} {IEEE Journal of Selected Topics in Quantum
  Electronics}\ }\textbf {\bibinfo {volume} {18}},\ \bibinfo {pages} {1763}
  (\bibinfo {year} {2012})}\BibitemShut {NoStop}%
\bibitem [{\citenamefont {Kelf}\ \emph {et~al.}(2005)\citenamefont {Kelf},
  \citenamefont {Sugawara}, \citenamefont {Baumberg}, \citenamefont
  {Abdelsalam},\ and\ \citenamefont {Bartlett}}]{Kelf_2005}%
  \BibitemOpen
  \bibfield  {author} {\bibinfo {author} {\bibfnamefont {T.~A.}\ \bibnamefont
  {Kelf}}, \bibinfo {author} {\bibfnamefont {Y.}~\bibnamefont {Sugawara}},
  \bibinfo {author} {\bibfnamefont {J.~J.}\ \bibnamefont {Baumberg}}, \bibinfo
  {author} {\bibfnamefont {M.}~\bibnamefont {Abdelsalam}},\ and\ \bibinfo
  {author} {\bibfnamefont {P.~N.}\ \bibnamefont {Bartlett}},\ }\bibfield
  {title} {\bibinfo {title} {Plasmonic band gaps and trapped plasmons on
  nanostructured metal surfaces},\ }\href
  {https://doi.org/10.1103/PhysRevLett.95.116802} {\bibfield  {journal}
  {\bibinfo  {journal} {Phys. Rev. Lett.}\ }\textbf {\bibinfo {volume} {95}},\
  \bibinfo {pages} {116802} (\bibinfo {year} {2005})}\BibitemShut {NoStop}%
\end{thebibliography}
%

\end{document}